\documentclass[prb,aps,superscriptaddress,twocolumn,nobibnotes]{revtex4-1}

\usepackage{graphicx} 
\usepackage{amsmath}
\usepackage{txfonts}
\usepackage[x11names]{xcolor}
\usepackage[hidelinks]{hyperref}
\usepackage{overpic}
\usepackage{booktabs}
\usepackage{soul}

\allowdisplaybreaks

\hypersetup{
	colorlinks = true, 
	urlcolor   = blue, 
	linkcolor  = blue, 
	citecolor  = blue  
}

\makeatletter
\def\p@subsection{}
\makeatother

\newcommand{\sifig}[1]{Figure~S{#1}}
\newcommand{\sitab}[1]{Table~S{#1}}

\newcommand{\Samsung}{Innovation Center, Samsung Electronics, Hwaseong 18448, Republic of Korea.}
\newcommand{\SAIT}{Samsung Advanced Institute of Technology, Samsung Electronics, Suwon 16678, Republic of Korea.}
\newcommand{\MIT}{Department of Chemistry, Massachusetts Institute of Technology (MIT), Cambridge, MA 02139, U.S.A.}
\newcommand{\KAIST}{Department of Chemistry, Korea Advanced Institute of Science and Technology (KAIST), Daejeon 34141, Republic of Korea.}

\newcommand{\sing}[1]{S\textsubscript{#1}}

\begin{document}

\title{Very-Large-Scale GPU-Accelerated Nuclear Gradient of Time-Dependent Density Functional Theory
	with Tamm--Dancoff Approximation and Range-Separated Hybrid Functionals}

\author{Inkoo~Kim} \affiliation{\Samsung} \affiliation{\MIT}
\author{Daun~Jeong} \affiliation{\Samsung}
\author{Leah~Weisburn} \affiliation{\MIT}
\author{Alexandra~Alexiu} \affiliation{\MIT}
\author{Troy~Van~Voorhis} \email{tvan@mit.edu} \affiliation{\MIT}
\author{Young~Min~Rhee} \email{ymrhee@kaist.ac.kr} \affiliation{\KAIST}
\author{Won-Joon~Son} \email{wonjoon.son@samsung.com} \affiliation{\Samsung}
\author{Hyung-Jin~Kim} \email{hj.windy.kim@samsung.com} \affiliation{\Samsung}
\author{Jinkyu~Yim} \affiliation{\Samsung}
\author{Sungmin~Kim} \affiliation{\SAIT}
\author{Yeonchoo~Cho} \affiliation{\SAIT}
\author{Inkook~Jang} \affiliation{\Samsung}
\author{Seungmin~Lee} \affiliation{\Samsung}
\author{Dae~Sin~Kim} \affiliation{\Samsung}

\begin{abstract}
Modern graphics processing units (GPUs) provide an unprecedented level of computing power.
In this study, we present a high-performance, multi-GPU implementation 
of the analytical nuclear gradient for
Kohn--Sham time-dependent density functional theory (TDDFT),
employing the Tamm--Dancoff approximation (TDA) and Gaussian-type 
atomic orbitals as basis functions.
We discuss GPU-efficient algorithms for the derivatives of electron repulsion integrals 
and exchange--correlation functionals 
within the range-separated scheme.
As an illustrative example, we calculated the TDA-TDDFT gradient of the \sing{1} state of
a full-scale green fluorescent protein with explicit water solvent molecules,
totaling 4353 atoms, at the $\omega$B97X/def2-SVP level of theory.
Our algorithm demonstrates favorable parallel efficiencies
on a high-speed distributed system equipped with 256 Nvidia A100 GPUs,
achieving $>$70\% with up to 64 GPUs and 31\% with 256 GPUs,
effectively leveraging the capabilities of modern high-performance computing systems.
\end{abstract}

\maketitle

\section{Introduction}

High-performance computing is undergoing a rapid evolution
supported by high-speed distributed systems that incorporate
heterogenous accelerators such as graphics processing units (GPUs)
for high-throughput data-parallel computation.\cite{Bourzac2017,McInnes2021,Heldens2021}
This heterogeneous architecture poses a challenge 
in efficiently adapting existing algorithms to fully exploit
the computational potential of these accelerators.
Within general-purpose GPU programming models,\cite{Herten2023}
fine-grained computational workloads represented by threads 
are organized into thread-blocks and are offloaded to GPU,
which consists of a large number of lightweight processors 
for massively parallel execution.
Therefore, a key consideration in achieving high performance 
with GPU acceleration is maintaining a high degree of concurrency
by simultaneously handling a large number of similar tasks
to utilize the single instruction multiple-thread (SIMT)-based GPU architecture effectively.

To harness the high-throughput potential of GPUs in quantum chemistry,
significant efforts have been made to redesign
conventional algorithms and parallelization tactics.
Kohn--Sham density functional theory (DFT),\cite{Kohn1965}
arguably the most widely adopted quantum chemistry method
for its balance of accuracy and cost-efficiency,\cite{
	Burke2012,Becke2014,Bursch2022}
has been a prime target for GPU acceleration.\cite{Gordon2020a,Felice2023,Vallejo2023a}
Focusing on DFT in the linear combination of atomic orbitals (LCAO) framework, 
using atom-centered Gaussian basis sets is a natural choice for molecular systems.\cite{Nagy2017}
In this approach, it is well known that
owing to the evaluation of four-index electron repulsion integrals (ERIs),
the integral-direct Fock build 
is the most computationally intensive section.\cite{Almlof1982}

Early pioneering works by Yasuda\cite{Yasuda2008} and by
Ufimtsev and Mart\'inez\cite{Ufimtsev2008} 
focused on accelerating the computation of ERIs
for the Hartree--Fock (HF) method on GPUs.
These were followed by numerous reports on 
improvements in both performance and scalability
to date.\cite{
	Ufimtsev2009a,Asadchev2010,Titov2013,
	Miao2013,Miao2015,Kussmann2017,
	Rak2015,Tornai2019,Barca2020,
	Qi2023,Vallejo2023,Asadchev2023a}
Recent advancements included 
efficient ground-state DFT calculations on multi-GPUs,
utilizing the GPU implementations for the numerical integration of 
exchange--correlation (XC) functionals,\cite{
	Seritan2020,Manathunga2021,Barca2021,Manathunga2020,Williams2020}
which allowed massively parallel simulations 
of very large molecules.\cite{Kowalski2021,Johnson2022}
Notably, scaling-reducing schemes such as
the resolution-of-identity\cite{Kussman2021,Asadchev2023,Williams2023}
and fragmentation-based methods\cite{Vallejo2023a} have also been 
explored on GPUs.
The electronic structures of excited states play crucial roles
from both chemical and materials perspectives
and can be modeled using 
the time-dependent variant of DFT (TDDFT).\cite{Runge1984,Dreuw2005}
Despite the widespread adoption of TDDFT across various fields,\cite{Herbert2022}
few studies have focused on its implementation on GPUs.\cite{Isborn2011,Kim2023}

Inspired by the GPU acceleration of TDDFT by Isborn et al.,\cite{Isborn2011}
some of present authors have provided in ref~\citenum{Kim2023}
a modernized implementation utilizing
the state-of-the-art Nvidia A100 GPUs
over high-speed Infiniband HDR networks 
for very large-scale calculations.
In this process, the global hybrid functionals 
showed limitations in excited-state calculations of 
full-scale biological proteins, 
as the occurrence of spurious low-lying charge-transfer states 
compromised their practical application.
As a simple, pragmatic remedy, we employed 
range-separated hybrid (RSH) functionals\cite{Leininger1997,Vydrov2006}
to address certain fundamental issues with global hybrid functionals.
Further, even with well-established 
parallel CPU implementations of the TDDFT nuclear gradient
in many quantum chemistry programs,\cite{
	Epifanovsky2021,Franzke2023,Zahariev2023,Mejia2023,Neese2022,Cao2021,Rinkevicius2020,g16}
to our knowledge, the GPU counterpart
in conjunction with RSH functionals,
which is critical for geometry optimizations of large systems
to find stationary points on the excited-state potential energy surface
and for elucidating various response properties,
has not been reported thus far.

In this work, as a necessary advance towards
reliable quantum chemistry calculations of excited states on an unprecedented scale,
we present a massively parallel GPU implementation of 
the nuclear gradient of the TDDFT method with RSH functionals,
employing the Gaussian-type atomic orbitals
for exploring the excited states of very large molecular systems.
We discuss efficient multi-GPU algorithms 
for computing the derivatives of ERIs and XC functionals,
and demonstrate strong scalability in the nuclear gradient calculations
of the singlet excited state of a 4353-atom protein system.
This implementation achieves 
reasonably high parallel efficiencies up to 256 Nvidia A100 GPUs.

\section{Theory and Algorithm}

\subsection{TDA-TDDFT Nuclear Gradient with RSH Functionals}

Within the linear-response formalism,
the total excited-state energy, $E$, of TDDFT is calculated as the sum of 
the ground-state energy, $E_0$, and the excitation energy, $E_\Delta$.
The Tamm--Dancoff approximation (TDA)\cite{Hirata1999} simplifies
the calculation of $E_\Delta$ by 
neglecting the deexcitation term in the Casida equation.\cite{Casida1995}
This approach allows $E_\Delta$ to be determined from the following eigenvalue equation:
\begin{align}
	\label{eq:tda}
	\mathbf{H} \mathbf{X} = E_\Delta \mathbf{X}
\end{align}
where $\mathbf{H}$ is the Hamiltonian represented within 
the space of (occupied--virtual) singly-excited Slater determinants
and $\mathbf{X}$ is the excitation amplitude
arranged in an $N_\mathrm{o} \times N_\mathrm{v}$ matrix,
where $N_\mathrm{o}$ and $N_\mathrm{v}$ refer to the numbers of 
occupied and virtual orbitals, respectively.
For simplicity, we limit our discussion to closed-shell orbitals
with restricted spins, using $i$, $j$ for the occupied orbitals, 
and $a$, $b$ for the virtual orbitals.
Throughout this work, the XC functionals considered are within 
the generalized gradient approximation (GGA).
These functionals depend on the density and its gradient,
and are described by
$f(\rho_\alpha, \rho_\beta, \gamma_{\alpha\alpha}, \gamma_{\alpha\beta},\gamma_{\beta\beta})$ 
where $\rho_\alpha$ and $\rho_\beta$ are the spin densities, and
$\gamma_{\alpha\alpha} \equiv \nabla \rho_\alpha {\cdot} \nabla \rho_\alpha$,
$\gamma_{\alpha\beta} \equiv \nabla \rho_\alpha {\cdot} \nabla \rho_\beta$ and
$\gamma_{\beta\beta} \equiv \nabla \rho_\beta {\cdot} \nabla \rho_\beta$
represent the gradient invariants.\cite{Perdew1996}
Additionally, we focus on the spin-adapted formulation for singlet excited states,
noting that the triplet counterpart can be easily derived
as described in ref~\citenum{Bauernschmitt1996}. 

By expanding all the terms in eq~\ref{eq:tda},
the nuclear derivative of $E$ with respect to
an atomic center $\mathbf{A} \equiv \{ A_x, A_y, A_z \}$ 
of the $A$-th atom can be expressed as
\begin{align}
	\label{eq:dE_mo}
	\nabla_{\mathbf{A}} E  &= \nabla_{\mathbf{A}} E_0 
	+ \sum_{ia} \sum_{jb} X_{ia} X_{jb} \nabla_{\mathbf{A}} 
	[ \varepsilon_{a} \delta_{ab} - \varepsilon_{i} \delta_{ij} \\ \nonumber
	&+ 2(ia|jb) - \alpha (ij|ab) - \beta (ij|ab)_\mathrm{lr} + (ia| \hat{f}_\mathrm{XC} | jb) ]
\end{align}
where $\varepsilon_i$ and $\varepsilon_a$ represent the orbital energies of 
the occupied and virtual molecular orbitals, 
$\psi_i (\mathbf{r})$ and $\psi_a (\mathbf{r})$, respectively, and
$(ia|jb)$, $(ij|ab)$ and $(ij|ab)_\mathrm{lr}$ 
denote the Coulomb, HF exchange and long-range HF exchange integrals, respectively,
defined by
\begin{align}
	\label{eq:eri_j}
	(ia|jb) &=
	\iint \! \psi_i (\mathbf{r}) \psi_a (\mathbf{r}) \frac{1}{r}
	\psi_j (\mathbf{r}') \psi_b (\mathbf{r}')
	\,\mathrm{d}\mathbf{r} \mathrm{d}\mathbf{r}'
	\\
	\label{eq:eri_k}
	(ij|ab) &=
	\iint \! \psi_i (\mathbf{r}) \psi_j (\mathbf{r}) 
	\frac{1}{r}
	\psi_a (\mathbf{r}') \psi_b (\mathbf{r}')
	\,\mathrm{d}\mathbf{r} \mathrm{d}\mathbf{r}'
	\\
	\label{eq:lreri_k}
	(ij|ab)_\mathrm{lr} &=
	\iint \! \psi_i (\mathbf{r}) \psi_j (\mathbf{r}) 
	\frac{\operatorname{erf} (\omega r)}{r}
	\psi_a (\mathbf{r}') \psi_b (\mathbf{r}')
	\,\mathrm{d}\mathbf{r} \mathrm{d}\mathbf{r}'
\end{align}
with $r \equiv | \mathbf{r} - \mathbf{r}' |$.
Furthermore, $(ia|\hat{f}_\mathrm{XC}|jb)$ represents the singlet XC kernel defined by\cite{Bauernschmitt1996}
\begin{align}
	\label{eq:fxc}
	&(ia|\hat{f}_\mathrm{XC}|jb) = 
	\nonumber
	\\
	&\iint \Bigg[
	\left(
	\frac{\partial^2 f}{\partial \rho_\alpha^2} +
	\frac{\partial^2 f}{\partial \rho_\alpha \partial \rho_\beta}
	\right) 
	\psi_i (\mathbf{r}) \psi_a (\mathbf{r}) \psi_j (\mathbf{r}') \psi_b (\mathbf{r}') \nonumber
	\\
	&+ \left(
	2\frac{\partial f}{\partial \gamma_{\alpha\alpha}} +
	\frac{\partial f}{\partial \gamma_{\alpha\beta}} 
	\right) 
	\nabla \big( \psi_i (\mathbf{r}) \psi_a (\mathbf{r}) \big) \cdot
	\nabla \big( \psi_j (\mathbf{r}') \psi_b (\mathbf{r}') \big) \nonumber
	\\
	&+ \left(
	\frac{\partial^2 f}{\partial \rho_\alpha \partial \gamma_{\alpha\alpha}} +
	\frac{\partial^2 f}{\partial \rho_\alpha \partial \gamma_{\beta\beta}} +
	\frac{\partial^2 f}{\partial \rho_\alpha \partial \gamma_{\alpha\beta}}
	\right)
	\Big( \psi_i (\mathbf{r}) \psi_a (\mathbf{r}) \nonumber
	\\
	&\times \nabla \rho \cdot \nabla \big( \psi_j (\mathbf{r}') \psi_b (\mathbf{r}') \big) +
	\nabla \rho \cdot \nabla \big( \psi_i (\mathbf{r}) \psi_a (\mathbf{r}) \big) \psi_j (\mathbf{r}') \psi_b (\mathbf{r}') 
	\Big)
	\nonumber
	\\
	&+ \left(
	\frac{\partial^2 f}{\partial \gamma_{\alpha\alpha}^2} +
	\frac{1}{2} \frac{\partial^2 f}{\partial \gamma_{\alpha\beta}^2} +
	2 \frac{\partial^2 f}{\partial \gamma_{\alpha\alpha} \gamma_{\alpha\beta}} +
	\frac{\partial^2 f}{\partial \gamma_{\alpha\alpha} \gamma_{\beta\beta}}
	\right) \nonumber
	\\
	&\times \nabla \rho \nabla \big( \psi_i (\mathbf{r}) \psi_a (\mathbf{r}) \big) \cdot
	\nabla \rho \nabla \big( \psi_j (\mathbf{r}') \psi_b (\mathbf{r}') \big) 
	\Bigg] \,\mathrm{d}\mathbf{r} \mathrm{d}\mathbf{r}'
\end{align}
with the total density $\rho \equiv \rho_\alpha + \rho_\beta$.
The range-separated scheme blends
the short-range exchange DFT functional
and the long-range HF exchange using the error function ($\operatorname{erf}$),
with the parameter $\omega$ setting the cross-fading point.
In the limits of short-range ($r \rightarrow 0$)
and long-range ($r \rightarrow \infty$),
the portions of HF exchange are determined by $\alpha$ and $\alpha + \beta$, respectively.
For global hybrid functionals,
the HF exchange is kept at constant $\alpha$ by setting $\beta = 0$.

In the following discussion, 
to derive the atomic orbital formulation of eq~\ref{eq:dE_mo}
for computational efficiency,
we utilize the direct differentiation approach.\cite{
	Foresman1992,Caillie1999,Caillie2000,Shroll2004,Liu2010}
Alternatively, one can achieve the same formulation 
through the Lagrangian approach.\cite{Furche2002,Chiba2006} 
Despite rather involving algebraic derivation, 
the resulting expression is relatively concise and can be written in
matrix notation, facilitating efficient computational processing.

We rewrite eq~\ref{eq:dE_mo} in the atomic orbital basis $\{\phi_\mu\}$ as
\begin{align}
	\label{eq:dE_ao}
	\nabla_{\mathbf{A}} E &= 
	\sum_{\mu \nu} P_{\mu \nu} \nabla_\mathbf{A} H^\mathrm{core}_{\mu \nu}
	+ \sum_{\mu \nu} W_{\mu \nu} \nabla_\mathbf{A} S_{\mu \nu} 
	+ \nabla_\mathbf{A} V_\mathrm{nuc}
	\nonumber
	\\
	&+\sum_{\mu \nu \lambda \sigma}  
	\Gamma_{\mu \nu \lambda \sigma} \nabla_{\mathbf{A}} (\mu \nu | \lambda \sigma) 
	+\sum_{\mu \nu \lambda \sigma}  
	\Gamma_{\mu \nu \lambda \sigma}^{\,\mathrm{lr}} \nabla_{\mathbf{A}} (\mu \nu | \lambda \sigma)_\mathrm{lr} 
	\nonumber
	\\ 
	&+ \sum_{\mu\nu} P_{\mu\nu} \nabla_{\mathbf{A}} F^\mathrm{XC}_{\mu\nu}
	+ \sum_{\mu\nu\lambda\sigma} R_{\mu\nu} R_{\lambda\sigma} 
	\nabla_{\mathbf{A}} 
	(\mu \nu | \hat{f}_\mathrm{XC} | \lambda \sigma)
\end{align}
where $\mathbf{H}_\mathrm{core}$ represents the core Hamiltonian matrix,
$\mathbf{S}$ represents the overlap matrix, 
$\mathbf{F}_\mathrm{XC}$ represents the Kohn-Sham Fock matrix,\cite{Pople1992}
$(\mu \nu | \lambda \sigma)$ and $(\mu \nu | \lambda \sigma)_\mathrm{lr}$ denote
the ERIs from eqs~\ref{eq:eri_j} and \ref{eq:eri_k}, respectively, and
$V_\text{nuc}$ denotes the nuclear repulsion energy.
The total density matrix is defined by
\begin{align}
	\mathbf{P} &= \mathbf{P}_0 + \mathbf{P}_\Delta
\end{align}
with
\begin{align}
	\mathbf{P}_0 &= 2\mathbf{C}_\text{o}\mathbf{C}_\text{o}^T
	\\
	\label{eq:p_del}
	\mathbf{P}_\Delta &= 
	\widetilde{\mathbf{P}}_\Delta + 2 \mathbf{C}_\text{v} \mathbf{Z} \mathbf{C}_\text{o}^T
	\\
	\widetilde{\mathbf{P}}_\Delta &=
	\mathbf{C}_\text{v} \mathbf{X}^T \mathbf{X} \mathbf{C}_\text{v}^T
	- \mathbf{C}_\text{o} \mathbf{X} \mathbf{X}^T \mathbf{C}_\text{o}^T
\end{align}
where $\mathbf{P}_0$ denotes the ground-state density matrix,
$\mathbf{P}_\Delta$ and $\widetilde{\mathbf{P}}_\Delta$ denote
the relaxed and unrelaxed difference density matrices, respectively, 
and $\mathbf{Z}$ describes the virtual--occupied orbital relaxation
that can be obtained by solving a set of coupled-perturbed equations 
described in section~\ref{sec:cp}.
The transition density matrix is expressed as
\begin{align}
	\mathbf{R} = \mathbf{C}_\text{o} \mathbf{X} \mathbf{C}_\text{v}^T
\end{align}
with the molecular orbital coefficient matrix 
$\mathbf{C} \equiv \left[ \mathbf{C}_\mathrm{o}, \mathbf{C}_\mathrm{v}\right]$,
and we write the product of densities as
\begin{align}
	\label{eq:gamma}
	\Gamma_{\mu\nu\lambda\sigma} &= 
	2 R_{\nu\mu} R_{\lambda\sigma}
	- \alpha R_{\sigma\mu} R_{\lambda\nu}
	\nonumber
	\\
	&+ \frac{1}{2} \left((P_{\mu\nu} + 2 P^{\,\Delta}_{\mu\nu}) P_{\lambda\sigma}
	- \frac{\alpha}{2} (P_{\mu\sigma} + 2 P^{\,\Delta}_{\mu\sigma}) P_{\lambda\nu} \right)
	\\
	\Gamma_{\mu\nu\lambda\sigma}^{\,\mathrm{lr}} &= 
	- \beta R_{\sigma\mu} R_{\lambda\nu}
	- \frac{\beta}{4} (P_{\mu\sigma} + 2 P^{\,\Delta}_{\mu\sigma}) P_{\lambda\nu} 
\end{align}

The atomic orbital formulation of TDDFT nuclear gradient in eq~\ref{eq:dE_ao}
facilitates the integral-direct implementation,
enabling the calculation of the four-index ERIs 
as needed during computation
to avoid the heavy $\mathcal{O}(N^4)$ scaling in memory.
We provide the working expressions for the energy-weighted density matrix, $\mathbf{W}$,
ERI derivatives (fourth and fifth terms in eq~\ref{eq:dE_ao}) and 
XC derivatives (sixth and seventh terms in eq~\ref{eq:dE_ao})
in sections~\ref{sec:cp}, \ref{sec:eri} and \ref{sec:xc}, respectively.

\subsection{Coupled-Perturbed Kohn--Sham Equations} \label{sec:cp} 

The virtual--occupied orbital relaxation, $\mathbf{Z}$, 
representing the first-order derivative of molecular orbital coefficients,
is obtained by solving the coupled-perturbed equation\cite{Gerratt1968}
using the Z-vector method:\cite{Handy1984}
\begin{align}
	\label{eq:cphf}
	(\mathbf{d} - \mathbf{A}) \mathbf{Z} = \mathbf{L}
\end{align}
where the matrix elements on the left-hand side are given as
\begin{align}
	d_{ai} &= \varepsilon_i - \varepsilon_a
	\\
	\label{eq:cphf_a}
	A_{ai,bj} &= 4(ia|jb) - \alpha \Big[ (ij|ab) + (ib|aj) \Big]
	\nonumber
	\\
	&- \beta \Big[ (ij|ab)_\mathrm{lr} + (ib|aj)_\mathrm{lr} \Big]
	+ 2(ia | \hat{f}_\mathrm{XC} | jb)
\end{align}
and the Lagrangian vector on the right-hand side is given as
\begin{align}
	\mathbf{L} &= \mathbf{C}_\text{v}^T \mathbf{F}_{\widetilde{\mathbf{P}}_\Delta} \mathbf{C}_\text{o}
	+ \mathbf{C}_\text{v}^T \mathbf{F}_{\mathbf{R}} \mathbf{C}_\text{v} \mathbf{X}^T
	- \mathbf{X}^T \mathbf{C}_\text{o}^T \mathbf{F}_{\mathbf{R}} \mathbf{C}_\text{o}
	+ \mathbf{C}_\text{v}^T \mathbf{K} \mathbf{C}_\text{o}
\end{align}
where the Fock-type matrix $\mathbf{F}_{\widetilde{\mathbf{P}}_\Delta}$ is defined by
\begin{align}
	\label{eq:G_2e}
	F_{\mu\nu}^{\,\widetilde{\mathbf{P}}_\Delta} &= \sum_{\lambda \sigma} \widetilde{P}^{\,\Delta}_{\sigma \lambda} 
	\Big[ 
	2 (\mu\nu|\lambda\sigma) - \alpha (\mu\lambda|\nu\sigma)
	- \beta (\mu\lambda|\nu\sigma)_\mathrm{lr}
	\nonumber
	\\ 
	&+(\mu \nu | \hat{f}_\mathrm{XC} | \lambda \sigma) 
	\Big]
\end{align}
through the contraction of $\widetilde{\mathbf{P}}_\Delta$, and
$\mathbf{F}_{\mathbf{R}}$ can be defined similarly. 

The XC hyperkernel involving higher-order functional derivatives is calculated as
\begin{align}
	K_{\mu\nu} 
	&= \iiint \phi_\mu (\mathbf{r}) \phi_\nu (\mathbf{r})
	\Big(  
	2 {\mathbf{h}}^2 f^{\,(2)}_2
	+ {g}^2 f^{\,(3)}_1
	+ 2 {g} (\nabla \rho \cdot {\mathbf{h}}) f^{\,(3)}_2
	\nonumber
	\\
	&+ (\nabla \rho \cdot {\mathbf{h}})^2 f^{\,(3)}_3
	\Big)
	+ \nabla \rho \cdot \nabla \big( \phi_\mu (\mathbf{r}) \phi_\nu (\mathbf{r}) \big) 
	\Big(
	2 {\mathbf{h}}^2 f^{\,(2)}_3
	+ {g}^2 f^{\,(3)}_2
	\nonumber
	\\
	& + 2 {g} (\nabla \rho \cdot {\mathbf{h}}) f^{\,(3)}_3
	+ (\nabla \rho \cdot {\mathbf{h}})^2 f^{\,(3)}_4
	\Big) \,\mathrm{d}\mathbf{r} \mathrm{d}\mathbf{r}' \mathrm{d}\mathbf{r}''
\end{align}
with the intermediates defined by
\begin{align}
	\label{eq:gt}
	{g} &= \sum_{\mu \nu} {R}_{\mu\nu} \phi_\mu (\mathbf{r}') \phi_\nu (\mathbf{r}'')
	\\
	\label{eq:ht}
	{\mathbf{h}} &= \sum_{\mu \nu} {R}_{\mu\nu} 
	\nabla \big( \phi_\mu (\mathbf{r}') \phi_\nu (\mathbf{r}'') \big)
\end{align}
and the second- and third-order functional derivatives organized as
\begin{align}
	f^{\,(2)}_2 &= \frac{\partial^2 f}{\partial \rho_\alpha \partial \gamma_{\alpha\alpha}}
	+ \frac{\partial^2 f}{\partial \rho_\alpha \partial \gamma_{\alpha\beta}}
	+ \frac{\partial^2 f}{\partial \rho_\alpha \partial \gamma_{\beta\beta}}
	\\
	f^{\,(2)}_3 &= \frac{\partial^2 f}{\partial \gamma_{\alpha\alpha}^2}
	+ \frac{1}{2} \frac{\partial^2 f}{\partial \gamma_{\alpha\beta}^2}
	+ 2 \frac{\partial^2 f}{\partial \gamma_{\alpha\alpha} \partial \gamma_{\alpha\beta}}
	+ \frac{\partial^2 f}{\partial \gamma_{\alpha\alpha} \partial \gamma_{\beta\beta}}
	\\
	f^{\,(3)}_1 &= \frac{\partial^3 f}{\partial \rho_\alpha^3}
	+ 3 \frac{\partial^3 f}{\partial \rho_\alpha^2 \partial \rho_\beta}
	\\
	f^{\,(3)}_2 &= \frac{\partial^3 f}{\partial \rho_\alpha^2 \partial \gamma_{\alpha\alpha}}
	+ \frac{\partial^3 f}{\partial \rho_\alpha^2 \partial \gamma_{\alpha\beta}}
	+ \frac{\partial^3 f}{\partial \rho_\alpha^2 \partial \gamma_{\beta\beta}}
	+ 2\frac{\partial^3 f}{\partial \rho_\alpha \partial \rho_\beta \partial \gamma_{\alpha\alpha}}
	\nonumber
	\\
	&+ \frac{\partial^3 f}{\partial \rho_\alpha \partial \rho_\beta \partial \gamma_{\alpha\beta}}
	\\
	f^{\,(3)}_3 &= \frac{\partial^3 f}{\partial \rho_\alpha \partial \gamma_{\alpha\alpha}^2}
	+ \frac{\partial^3 f}{\partial \rho_\beta \partial \gamma_{\alpha\alpha}^2}
	+ \frac{\partial^3 f}{\partial \rho_\alpha \partial \gamma_{\alpha\beta}^2}
	+ 2 \frac{\partial^3 f}{\partial \rho_\alpha \partial \gamma_{\alpha\alpha} \partial \gamma_{\alpha\beta}}
	\nonumber
	\\
	&+ 2 \frac{\partial^3 f}{\partial \rho_\beta \partial \gamma_{\alpha\alpha} \partial \gamma_{\alpha\beta}}
	+ 2 \frac{\partial^3 f}{\partial \rho_\alpha \partial \gamma_{\alpha\alpha} \partial \gamma_{\beta\beta}}
	\\
	f^{\,(3)}_4 &= \frac{\partial^3 f}{\partial \gamma_{\alpha\alpha}^3}
	+ \frac{1}{2} \frac{\partial^3 f}{\partial \gamma_{\alpha\beta}^3}
	+ 3 \frac{\partial^3 f}{\partial \gamma_{\alpha\alpha}^2 \partial \gamma_{\alpha\beta}}
	+ 3 \frac{\partial^3 f}{\partial \gamma_{\alpha\alpha}^2 \partial \gamma_{\beta\beta}}
	\nonumber
	\\
	&+ 3 \frac{\partial^3 f}{\partial \gamma_{\alpha\beta}^2 \partial \gamma_{\alpha\alpha}}
	+ 3 \frac{\partial^3 f}{\partial \gamma_{\alpha\alpha} \partial \gamma_{\alpha\beta} \partial \gamma_{\beta\beta}}
\end{align}
This can be implemented straightforwardly
by repurposing the existing routine for the XC kernel
used in the TDDFT energy computation.

To solve eq~\ref{eq:cphf} without the need to store $\mathbf{A}$ of large dimension
$N_\mathrm{o}N_\mathrm{v} \times N_\mathrm{o}N_\mathrm{v}$,
we employ the Krylov subspace method.
In this approach,
the orthogonal subspace $\mathbf{b} \equiv \{ \mathbf{b}_0, \mathbf{b}_1, \cdots \}$ 
is iteratively augmented to span the basis space for $\mathbf{Z}$
until the residual is acceptably small.\cite{Pople1979}
We rewrite eq~\ref{eq:cphf} to adapt it into the Krylov method as
\begin{align}
	(\mathbf{I} - \mathbf{A}') \mathbf{Z} = \mathbf{b}_0
\end{align}
where $\mathbf{A}' = \mathbf{d}^{-1}\mathbf{A}$ and
$\mathbf{b}_0 = \mathbf{d}^{-1}\mathbf{L}$.
At the $k$-th iteration,
the $(k{+}1)$-dimensional subspace linear equation is solved to obtain $\mathbf{Z}_k$:
\begin{align}
	\widetilde{\mathbf{A}}_k \mathbf{a}_k = \mathbf{x}_k
\end{align}
where the elements are defined as
\begin{align}
	\widetilde{A}_{i,j}^{\,(k)} &= \mathbf{b}_i^T \mathbf{b}_j - \mathbf{b}_i^T \widetilde{\boldsymbol{\sigma}}_j
	\\
	x_i^{\,(k)} &= \delta_{0i} \mathbf{b}_i^T \mathbf{b}_i
\end{align}
The scaled matrix-vector product
$\widetilde{\boldsymbol{\sigma}}_k \equiv \mathbf{d}^{-1} \mathbf{A}\mathbf{b}_k$ 
is calculated directly as in the Fock build in TDDFT:
\begin{align}
	\label{eq:sigma}
	\boldsymbol{\sigma}_k &= \mathbf{A}\mathbf{b}_k
	\\ \nonumber
	&= \mathbf{C}_\text{v}^T 
	\widetilde{\mathbf{F}}_k
	\mathbf{C}_\text{o}
\end{align}
Here, we use a slightly modified expression for the Fock-type matrix,
derived from eq~\ref{eq:cphf_a}, written as
\begin{align}
	\label{eq:zfock}
	\widetilde{F}_{\mu\nu}^{\,(k)}
	&= \sum_{\lambda \sigma} B_{\sigma \lambda}^{\,(k)}
	\Big[ 
	4(\mu\nu|\lambda\sigma) 
	- \alpha \,\Big( (\mu\lambda|\sigma\nu) + (\mu\sigma|\lambda\nu) \Big)
	\nonumber
	\\
	&- \beta \,\Big( (\mu\lambda|\sigma\nu)_\mathrm{lr} + (\mu\sigma|\lambda\nu)_\mathrm{lr}  \big)
	+ 2 (\mu \nu | \hat{f}_\mathrm{XC} | \lambda \sigma) 
	\Big]
\end{align}
where $\mathbf{B}_k \equiv \mathbf{C}_\text{v} \mathbf{b}_k \mathbf{C}_\text{o}^T$.

Then, $\mathbf{Z}$ can be approximated as
\begin{align}
	\mathbf{Z} &\simeq \mathbf{Z}_k = \sum_{i=0}^k a_i \mathbf{b}_i
\end{align}
If the residual is not sufficiently small,
an additional subspace vector is constructed,
\begin{align}
	\mathbf{b}_{k+1} &= \widetilde{\boldsymbol{\sigma}}_k - \sum_{i=0}^k 
	\frac{\mathbf{b}_i^T \widetilde{\boldsymbol{\sigma}}_k}{\mathbf{b}_i^T \mathbf{b}_i} \mathbf{b}_i
\end{align}
and the iteration continues.

Finally, the energy-weighted density matrix can be obtained through
matrix multiplications as
\begin{align}
	\label{eq:w_overlap}
	\mathbf{W} &= 
	\mathbf{C}_\mathrm{o} \operatorname{diag}(\boldsymbol{\varepsilon}_\mathrm{o}) 
	\,(2 \mathbf{I} - \mathbf{X}\mathbf{X}^T)\mathbf{C}_\mathrm{o}^T
	+ \mathbf{C}_\mathrm{v} \operatorname{diag}(\boldsymbol{\varepsilon}_\mathrm{v})
	\mathbf{X}^T \mathbf{X} \mathbf{C}_\mathrm{v}^T
	\nonumber
	\\
	&+ 2 \mathbf{C}_\mathrm{v} \mathbf{Z} \operatorname{diag}(\boldsymbol{\varepsilon}_\mathrm{o})
	\mathbf{C}_\mathrm{o}^T
	+ \frac{1}{4} \mathbf{P}_0^T ( \mathbf{F}_{\mathbf{P}_\Delta} + \mathbf{K} ) \mathbf{P}_0
	+ \frac{1}{2} \mathbf{P}_0 \mathbf{F}_{\mathbf{R}}^T \mathbf{R}^T
	\nonumber
	\\
	&+ \mathbf{C}_\mathrm{v} \mathbf{C}_\mathrm{v}^T \mathbf{F}_\mathbf{R}^T \mathbf{R}
	+ \mathbf{R}^T \mathbf{F}_\mathbf{R} \mathbf{P}_0
\end{align}
where $\boldsymbol{\varepsilon}_\mathrm{o}$ and $\boldsymbol{\varepsilon}_\mathrm{v}$
denote the orbital energies of the occupied and virtual molecular orbitals, respectively.

\subsection{ERI Derivatives and Their GPU-Acceleration} \label{sec:eri}

The Cartesian Gaussian function centered on the $A$-th atom is given by
\begin{align}
	\label{eq:cartG}
	\phi_\mu (\mathbf{r}; \zeta, \mathbf{l}, \mathbf{A}) &= N (x - A_x)^{l_x} (y - A_y)^{l_y} (z - A_z)^{l_z}
	\\ \nonumber
	&\times \exp ( -\zeta | \mathbf{r} - \mathbf{A} |^2 )
\end{align}
where $N$ denotes the normalization constant, 
$\mathbf{l} \equiv (l_x, l_y, l_z)$ represents the Cartesian quantum numbers, and
$\zeta$ denotes the exponent.
For the given angular momentum, $l \equiv l_x + l_y + l_z$,
there are $n_\mathrm{cart} = (l+1)(l+2)/2$ Cartesian functions
or $n_\mathrm{sph} = 2l+1$ spherical functions, 
the latter of which being easily be obtained via a spherical transformation of the former.
The derivative of a Cartesian Gaussian function can be expressed as
a linear combination of functions with higher and lower angular momentum.
For instance, the $x$-derivative can be written as
\begin{align}
	\label{eq:deriv}
	\nabla_x \phi_\mu^{l_x,l_y,l_z} = l_x \phi_\mu^{l_x-1,l_y,l_z} 
	- 2 \zeta \phi_\mu^{l_x+1,l_y,l_z}
\end{align}
Because the Gaussian functions are atom-centered,
the ERI gradient with respect to the $E$-th atom can be expressed as
\begin{align}
	\label{eq:deri}
	\nabla_\mathbf{E} (\mu_A \nu_B |\lambda_C \sigma_D) &=
	- \delta_{EA} (\mu'_A  \nu_B|\lambda_C \sigma_D) 
	- \delta_{EB} ( \mu_A \nu'_B|\lambda_C \sigma_D)
	\nonumber
	\\
	&- \delta_{EC} (\mu_A \nu_B|\lambda'_C  \sigma_D)
	 - \delta_{ED} (\mu_A \nu_B| \lambda_C \sigma'_D)
\end{align}
where the subscript denotes the atomic center, and
the prime denotes the Cartesian derivative
(i.e., $\mu'_A \equiv \nabla \phi_\mu $).
The derivatives of long-range ERIs can be defined similarly.
Moreover, one of the derivative terms in eq~\ref{eq:deri} can be obtained at no cost
by using translational invariance,
$\sum_E \nabla_\mathbf{E} (\mu_A \nu_B | \lambda_C \sigma_D) = 0 $;\cite{Komornicki1977}
for instance, when all the centers are different, we can write
\begin{align}
	\label{eq:ti}
	(\mu'_A \nu_B | \lambda_C \sigma_D) &= 
	-(\mu_A \nu'_B | \lambda_C \sigma_D)
	-(\mu_A \nu_B | \lambda'_C \sigma_D)
	\nonumber
	\\
	&- (\mu_A \nu_B | \lambda_C \sigma'_D)
\end{align}
This reduces the computational cost for ERI derivatives by at least one fourth
if $\phi_\mu$ was chosen for the highest angular momentum.
With the aid of permutation symmetry in ERI,
we placed the atomic orbital of the highest angular momentum 
at the first index, $\mu$, in our code to directly use the above equation.

In the integral-direct algorithm,
the contributions to nuclear gradients up to the four centers
comprising an ERI are calculated.
Focusing on the case where all the four centers are different
for simplicity, the update of $\nabla_\mathbf{A} E$ in eq~\ref{eq:dE_ao}
by the ERI derivatives
with the given $\phi_\mu$, $\phi_\nu$, $\phi_\lambda$ and $\phi_\sigma$
can be expressed as
\begin{align}
	\label{eq:eriupdate}
	\nabla_{A_x} E &\mathrel{{+}{=}} \Gamma_{\mu\nu\lambda\sigma}
	\left[
	2 \zeta (\mu_{A}^{+} \nu_{B} | \lambda_{C} \sigma_{D})
	- l_x (\mu_{A}^{-} \nu_{B} | \lambda_{C} \sigma_{D})
	\right]
	\nonumber
	\\
	&\,\,\,\,+ \Gamma_{\mu\nu\lambda\sigma}^{\,\mathrm{lr}}
	\left[
	2 \zeta (\mu_{A}^{+} \nu_{B} | \lambda_{C} \sigma_{D})_\mathrm{lr}
	- l_x (\mu_{A}^{-} \nu_{B} | \lambda_{C} \sigma_{D})_\mathrm{lr}
	\right]
\end{align}
where the superscripts $+$ and $-$ denote the raising and lowering of 
Cartesian quantum numbers, i.e., $(l_x \pm 1, l_y, l_z)$.
$\nabla_{A_y} E$ and $\nabla_{A_z} E$ can be expressed in an identical manner, and
similar expressions can be readily obtained for the other three centers.
Generally, $\phi_\mu$ is described by a set of Gaussian functions as
\begin{align}
	\phi_\mu (\mathbf{r}) = \sum_k^{K_\mu} d_k \varphi_k (\mathbf{r})
\end{align}
where $K_\mu$ denotes the degree of contraction, and $d_k$ denotes the coefficient.
Therefore, for a given ERI $(\mu\nu|\lambda\sigma)$,
nested loops over each contraction occur
with a total loop-count of $K_\mu {\cdot} K_\nu {\cdot} K_\lambda {\cdot} K_\sigma$,
each evaluating a batch of ERI derivatives with a dimension of
$3 \times n_\mu \times n_\nu \times n_\lambda \times n_\sigma$ in the code
where $n$ denotes $n_\mathrm{cart}$ or $n_\mathrm{sph}$.

To simultaneously achieve concurrency of GPU kernels and load-balancing 
within the distributed GPUs
while minimizing the thread divergence,
we have extended the multi-GPU ERI algorithm described in our previous report.\cite{Kim2023}
The list of non-zero shell-pairs is presorted
according to the degree of contractions, $K$, and the Schwarz upper-bound
such that when two lists are combined to form an ERI supermatrix,
each thread-blocks of optimal dimension will likely exhibit identical loop-counts
for maximum thread concurrency within the GPU kernel.
Block-cyclic distributions of the thread-blocks are used
to achieve the load-balance between distributed GPUs.
The non-redundant ERI types according to the sequence of the angular momenta in indices
are computed sequentially to mitigate the thread divergence.
For instance, there are a total of 21 GPU kernels each representing a specific ERI type
for the atomic orbitals with $l \le 2$.

\begin{figure}[t]
	\centering
		\includegraphics[width=8cm,trim={5.7cm 10.7cm 5.9cm 4.3cm},clip]{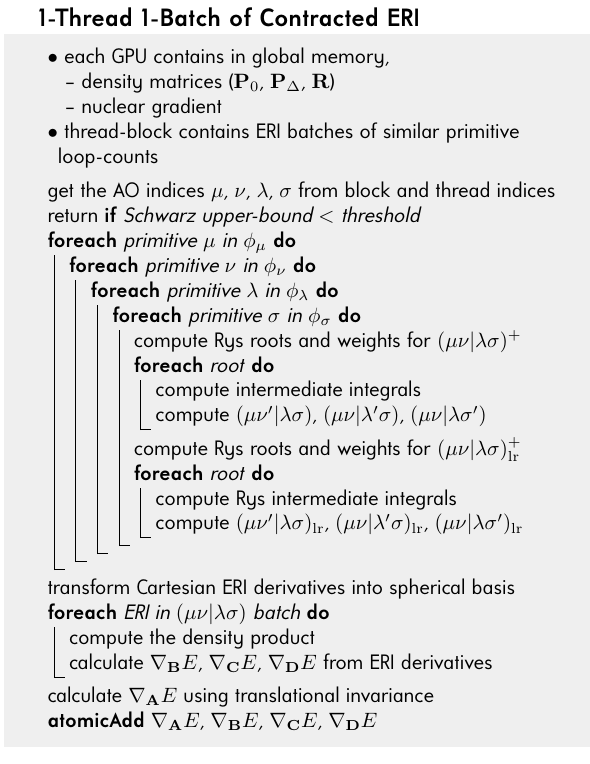}
	\caption{\normalfont
		Sketch of the algorithm used to compute the ERI derivative on GPU.}
	\label{fig:algo-eri}
\end{figure}

The GPU implementation of eq~\ref{eq:eriupdate} is shown in Figure~\ref{fig:algo-eri}.
For all ERI calculations, we employed
the Rys quadrature method\cite{Dupuis1976}
since it requires the least memory footprint 
compared to other algorithms based on the recursion formula.
The Rys roots and weights
that correspond to the higher angular momentum ERI
are reused for the ERI with lower angular momenta.
Thus, the number of Rys roots is given by
$n_\mathrm{rys} = (l_\mu + l_\nu + l_\lambda + l_\sigma + 1)/2 + 1$,
which is used for 
$(\mu\nu^+|\lambda\sigma)$, $(\mu\nu|\lambda^+\sigma)$, $(\mu\nu|\lambda\sigma^+)$,
$(\mu\nu^-|\lambda\sigma)$, $(\mu\nu|\lambda^-\sigma)$ and $(\mu\nu|\lambda\sigma^-)$.

While it is technically possible to reuse 
the ERI routines for higher angular momenta
to calculate the ERI derivatives,
such an implementation would introduce complexity 
in exploiting the 8-fold permutation symmetry
as well as utilizing the translational invariance,
which can effectively remove the most computationally intensive derivatives.
Therefore, we have explicitly constructed
independent GPU kernels for the ERI derivatives.

\subsection{Numerical Integration of XC Derivatives on GPUs} \label{sec:xc}

The XC functionals and their $n$-th order derivatives can be calculated by
numerical integration over well-defined quadrature grids as
\begin{align}
	\label{eq:xcquad}
	I = \sum_q w_q f^{\,(n)} (\mathbf{r}_q)
\end{align}
where $w_q$ denotes the grid weight at point $\mathbf{r}_q$.
In the case of polyatomic systems, atom-centered spherical grids can be constructed
by combining a radial grid and an angular grid.
We employed the Euler--Maclaurin grid\cite{Murray1993} for the radial partitioning
and the Lebedev grid\cite{Lebedev1977} for the angular partitioning.
The numerical stability of discrete summation can be maintained
by Becke's space partitioning method,\cite{Becke1988}
in which weights are obtained via the fuzzy cell construction.

For the given quadrature point at $\mathbf{r}_q$,
the updating expression for the nuclear gradient by
the XC contribution in eq~\ref{eq:dE_ao},
can be written as 
\begin{align}
	\label{eq:dxc}
	\nabla_{\mathbf{A}} E 
	&\mathrel{{+}{=}}
	- 2 w_q \sideset{}{'}\sum_\mu \sum_\nu 
	\bigg[
	P_{\mu\nu} 
	\Big( f^{\,(1)}_1 \mathbf{x} + \frac{1}{2} f^{\,(1)}_2 \mathbf{y} \Big)
	+ 2 R_{\mu\nu}
	\Big(
	f^{\,(1)}_2 \mathbf{z}	
	\nonumber
	\\
	&+f^{\,(2)}_1 {g} \mathbf{x}
	+ f^{\,(2)}_2
	\big( g \mathbf{y} + ({\mathbf{h}} \cdot \nabla \rho) \mathbf{x} \big)
	+ f^{\,(2)}_3 ({\mathbf{h}} \cdot \nabla \rho) \mathbf{y}
	\Big)
	\nonumber
	\\	
	&+ \frac{1}{2} P^{\,0}_{\mu\nu}
	\Big(
	    f^{\,(1)}_2   \mathbf{z}_\Delta
	+   f^{\,(2)}_1 g_\Delta \mathbf{x}
	+   f^{\,(2)}_2 \big( g_\Delta \mathbf{y}
	+   (\mathbf{h}_\Delta \cdot \nabla \rho ) \mathbf{x} 
	\nonumber
	\\
	&+2 \mathbf{h}^2 \mathbf{x}
	+ 4 g \mathbf{z} \big)
	+ f^{\,(2)}_3 \big( 
	2 \mathbf{h}^2 \mathbf{y} + ( \mathbf{h}_\Delta \cdot \nabla ) \rho \mathbf{y}
	+ 4 ( \mathbf{h} \cdot \nabla \rho ) \mathbf{z} 
	\big)
	\nonumber
	\\
	&+  f^{\,(3)}_{1} g^2 \mathbf{x}
	+   f^{\,(3)}_{2} \big( g^2 \mathbf{y}
	+ 2 g (\mathbf{h} \cdot \nabla \rho) \mathbf{x} \big)
	+   f^{\,(3)}_{3} \big( (\mathbf{h} \cdot \nabla \rho)^2 \mathbf{x}
	\nonumber
	\\
	&+2 g (\mathbf{h} \cdot \nabla \rho) \mathbf{y} \big)
	+   f^{\,(3)}_{4} (\mathbf{h} \cdot \nabla \rho)^2 \mathbf{y}
	\Big)
	\bigg]
\end{align}
where the primed sum ($\Sigma'$) indicates that the sum runs over the atomic orbitals centered on $A$.
The functional derivatives are defined as
\begin{align}
	f^{\,(1)}_1 &= \frac{\partial f}{\partial \rho_\alpha}
	\\
	f^{\,(1)}_2 &= 2 \frac{\partial f}{\partial \gamma_{\alpha\alpha}}
	+ \frac{\partial f}{\partial \gamma_{\alpha\beta}}
	\\
	f^{\,(2)}_1 &= \frac{\partial^2 f}{\partial \rho_\alpha^2}
	+ \frac{\partial^2 f}{\partial \rho_\alpha \partial \rho_\beta}
\end{align}
The intermediates contracting the transition density matrix are defined as
\begin{align}
	\label{eq:g}
	g_\Delta &= \sum_{\mu \nu} {P}^{\,\Delta}_{\mu\nu} \phi_\mu \phi_\nu 
	\\
	\label{eq:h}
	\mathbf{h}_\Delta &= \sum_{\mu \nu} {P}^{\,\Delta}_{\mu\nu} \nabla ( \phi_\mu \phi_\nu )
\end{align}
The recurring vectors are defined as
\begin{align}
	\mathbf{x} & = \nabla \phi_\mu \cdot \phi_\nu
	\\
	\mathbf{y} &= [ \phi_\nu \nabla (\nabla \phi_\mu)^T 
	+ \nabla \phi_\mu (\nabla \phi_\nu )^T ] \nabla \rho
	\\
	\mathbf{z} &= [ \phi_\nu \nabla (\nabla \phi_\mu)^T 
	+ \nabla \phi_\mu (\nabla \phi_\nu )^T ] {\mathbf{h}}
	\\
	\mathbf{z}_\Delta &= [ \phi_\nu \nabla (\nabla \phi_\mu)^T 
	+ \nabla \phi_\mu (\nabla \phi_\nu )^T ] \mathbf{h}_\Delta
\end{align}
We note that the weight derivative has been omitted
as it is often negligible in the case of nuclear gradients
when sufficiently fine grids are used.

\begin{figure}[t]
	\centering
		\includegraphics[width=8cm,trim={5.7cm 15.2cm 5.9cm 4.3cm},clip]{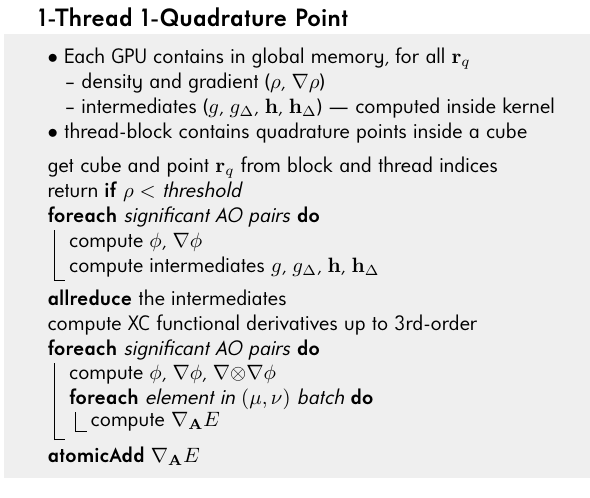}
	\caption{\normalfont
		Sketch of the algorithm used to compute the XC derivative on GPU}
	\label{fig:algo-xc}
\end{figure}

Figure~\ref{fig:algo-xc} shows the GPU implementation of the XC nuclear gradients.
Practically, owing to the locality of Gaussian functions
only the atomic orbitals in the vicinity of a quadrature point, 
which can be predetermined by a suitable gauging, 
are relevant in computing eq~\ref{eq:dxc}.
In other words, a sufficiently small cluster of quadrature points
will share most of these atomic orbitals.
Therefore, within GPU kernel, concurrency can be easily achieved by
progressively dividing the entire grid into octants until
all the points inside any octant can be assigned to
a one-dimensional thread-block.
By predetermining the atomic orbitals that contribute to the 
the thread-block (equivalently, a box containing the grid points in the vicinity),
we not only eliminate running the entire list of atomic orbitals, 
but more importantly
the threads can loop over the same list of significant atomic orbitals,
maximizing the concurrency.

Computational cost can be reduced by precomputing
atomic orbital-independent intermediates for each $\mathbf{r}_q$, 
${g}$ (eq~\ref{eq:gt}) and ${\mathbf{h}}$ (eq~\ref{eq:ht})
contracting $\mathbf{R}$, and
$g_\Delta$ (eq~\ref{eq:g}) and $\mathbf{h}_\Delta$ (eq~\ref{eq:h})
contracting ${\mathbf{P}}_{\Delta}$.
From the TDDFT energy calculations, $\rho$ and $\nabla \rho$
for all $\mathbf{r}_q$ can be stored and reused.
Therefore, the precomputed quantities require a total of 
12$n_q$ storage, 
where $n_q \equiv n_\mathrm{atom}{\cdot}n_\mathrm{rad}{\cdot}n_\mathrm{ang}$ 
is the number of quadrature points without grid pruning.
We provide selected cases in \sitab{2}, affirming that
the current GPU memory size is sufficient even for a very large molecular system.

\subsection{Implementation}

The GPU algorithm for the TDDFT nuclear gradient,
as presented in the previous sections, was implemented in an in-house code.
Our computational framework is based on core parts of the KPACK code,\cite{Kim2013}
a two-component relativistic quantum chemistry program,
and has been completely rewritten in Fortran 2003/2008
to accommodate the modern object-oriented features.
The GPU code has been written in CUDA Fortran\cite{cudaFor}
using the extensions to the Fortran language 
with several keywords and APIs to offload computations to GPUs.
To ensure efficient GPU utilization with XC functionals,
the C code of selected functionals from LibXC 5.2.3\cite{Lehtola2018}
has been converted into CUDA Fortran and integrated with our code.
We validated our implementation by comparing
analytic gradients with
numerical gradients obtained from the finite difference method.

To design an efficient parallel model for our distributed, 
high-performance GPU systems, as outlined in our previous report,\cite{Kim2023}
we adopted a hybrid MPI/OpenMP model.
We leveraged the message-passing interface (MPI) using OpenMPI 4.1.5 
with the multithreaded Unified communication-X (UCX) 1.12.1
for both point-to-point and one-sided communications between the nodes.
On each node, 
CPU-to-GPU bindings were achieved using OpenMP parallel interfaces.
Additionally, we utilized
the single-node multi-GPU linear algebra libraries of
Nvidia's cuBlasXt 11.11 and cuSolverMg 11.4 
for matrix multiplications and diagonalizations, respectively.
The Nvidia HPC SDK 22.11 compiler suite with the ``-fast'' optimization flag
was used for code compilation.

Our code employs a symmetric orthonormalization procedure 
with a threshold of $10^{-6}$~au for defining orthonormal atomic orbitals.
The SCF convergence threshold was set as the root-mean-squared (rms) difference 
of $10^{-6}$~au between the density matrices of two consecutive SCF iterations.
A density-multiplied Schwarz upper bound of $10^{-10}$~au was used as the threshold
for screening ERIs and their derivatives.
In numerical integration, a density threshold of $10^{-10}$~au 
for the grid point was applied.
Throughout this work, we used a (50, 194)-grid,
combining a 50-point radial grid with a 194-point angular grid per atom.
A residual threshold of $10^{-5}$~au was used in both 
Davidson diagonalization and Z-vector iteration.
Currently, all one-electron integrals and their derivatives 
are computed using CPUs over the hybrid OpenMP/MPI model.

\section{Application}

\begin{figure}[t]
	\centering
	\hspace*{-3cm}
	\begin{overpic}[height=6cm,trim={6cm 3cm 5cm 2.5cm}]{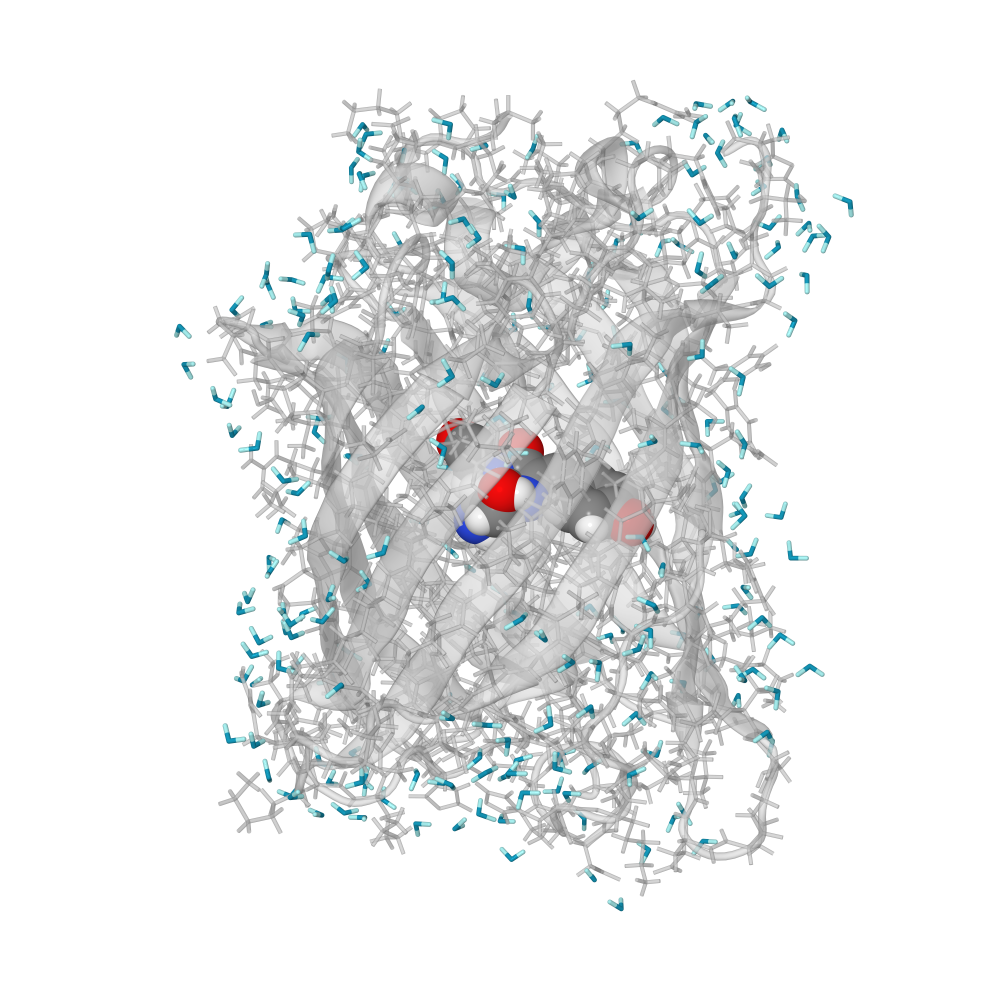}
		\put(74,32) {
			\includegraphics[height=2.4cm,trim={4.5cm 0cm 0cm 0cm},clip]{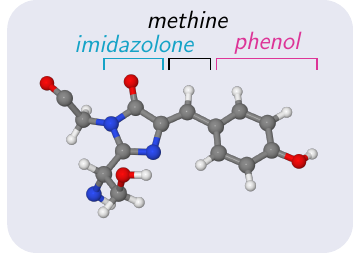}
		}
		\put(101, 34){\sffamily \normalsize HBI}
	\end{overpic}
	\caption{\normalfont
		Green fluorescent protein (GFP) in the presence of water solvent molecules. 
		The HBI chromophore is shown inside
		the protein cage, which is visually aided with ribbon representation.
	}
	\label{fig:gfp}
\end{figure}

In this section, we showcase the capability and performance of our TDDFT implementation
by calculating the excited-state nuclear gradient of a very large molecular system,
green fluoresecent protein (GFP).
GFP is a naturally occurring biological protein
extracted from \textit{Aequorea victoria},
and has become a commonly used tool in molecular and cell biology
for labeling and visualizing proteins in living cells.\cite{Zimmer2002}

As shown in Figure~\ref{fig:gfp},
within the rigid 11-stranded $\beta$-barrel comprising 238 amino acids,
the \textit{p}-hydroxybenzylidene-imidazolinone (HBI) chromophore
converts blue light to green light. 
The neutral and anionic forms of the chromphore are responsible 
for the light absorption in the ultraviolet and blue regions, respectively.
The HBI chromophore consists of
an imidazolone and a phenol moiety connected via a methine bridge.

In previous theoretical studies of GFP,
to characterize the relevant excited-state,
researchers utilized quantum chemical methods 
on the chromophore independently or with 
electrostatic/polarizable embedding of the surrounding environment via molecular mechanics
by the so-called QM/MM model.\cite{Laino2004,Epifanovsky2009,Nabo2017}
Of course, quantum chemical methods should outperform the empirical force field
when compared on an equal footing,\cite{Kulik2012}
and their usage could potentially benefit
as distant residues can also affect
the excitation energies and properties of the chromophore.\cite{Schwabe2015}
However, increasing the QM region is nontrivial,
requiring chemical intuitions of domain experts,
and tends to approach slowly to the accuracy of 
the large QM asymptotic limit.\cite{Kulik2016}

Moreover, when applied to large systems,
QM methods can experience challenges associated with the specific model chemistry.
TDDFT calculations with the pure or global hybrid functionals
often encounter low-lying spurious charge-transfer states 
stemming from an unphysically small band gap,
an artifact of the underestimated exchange energy at long distance,\cite{Rudberg2012}
and partially from unequilibrated electrostatic potential on surfaces.\cite{Lever2013}
Using RSH functionals with correct long-distance behavior
described by 100\% HF exchange is suitable for large systems,
as well as including solvent effect via implicit solvation models or 
explicit solvents, providing an effective solution
for reliable results from large-scale TDDFT calculation.

We employed the geometry of GFP prepared in ref~\citenum{Foresman2015} 
from the X-ray diffraction structure (PDB ID: 1W7S)\cite{VanThor2005}
with minor refinements to a few misplaced atoms.\cite{Kim2023}
This system represents a dense three-dimensional structure (Figure~\ref{fig:gfp}).
For calculations, the $\omega$B97X RSH GGA functional\cite{Chai2008} was chosen, 
as it has demonstrated to provide
reliable results over a wide range of interactions.\cite{Mardirossian2017}
The default parameters of $\omega$B97X were used:
$\alpha=0.158$, $\beta=0.842$, $\omega=0.3$~bohr$^{-1}$.
Explicit inclusion of water solvent molecules and 
utilization of the double-zeta plus polarization basis set, def2-SVP,\cite{Weigend2005}
afforded a more realistic model.
The total numbers of atoms and basis functions were 4353 and 40\,518, respectively,
representing the largest attempt to compute
the nuclear gradient of an excited state at the TDDFT level to our knowledge.

\subsection{Excited State of Green Fluorescent Protein}

Table~\ref{tbl:calc} presents the calculated results of the full-scale GFP system
at the TDA-TDDFT/$\omega$B97X level with def2-SVP basis sets,
alongside QM/MM results using the ONIOM model,\cite{Svensson1996}
aiming to unveil the long-distance quantum mechanical effect in the excited state.

\begin{table}[t]
	\caption{Calculated results of GFP at the TDA-TDDFT level}
	\label{tbl:calc}
	\begin{ruledtabular}
		\begin{tabular}{lcccc}
			& HOMO / LUMO & Band gap & \sing{1} & TDM \\
			method & (eV) & (eV) & (eV)\textsuperscript{\emph{a}} & (debye)\textsuperscript{\emph{b}} \\
			\hline
			ONIOM\textsuperscript{\emph{c}} & $-$3.21 / 4.20 & 7.41 & 3.80 & 7.74\\
			$\omega$B97X & $-$3.80 / 2.78 & 6.58 & 3.62 & 7.07\\
		\end{tabular}
	\end{ruledtabular}
	
	\raggedright
	\textsuperscript{\emph{a}} Vertical excitation energy.
	\textsuperscript{\emph{b}} Norm of transition dipole moment vector.
	\textsuperscript{\emph{c}} 
	The QM region containing the HBI chromophore (43 atoms) 
	was treated by the $\omega$B97X level,
	while the Amber force fields (ref~\citenum{Cornell1995})
	were used to describe the MM region.
	The QM/MM calculations were performed 
	using the Gaussian 16.C.01 program (ref~\citenum{g16}).
\end{table}

The orbital energies of the two approaches exhibit significant differences,
which can be attributed to 
the full-scale calculation representing the extended system.
What is more important is that the band gap of $\omega$B97X 
remains sufficiently large, rectifying the band gap underestimation 
observed with global hybrid functionals.
In a previous study, we obtained the band gap of 1.88~eV 
with B3LYP in the full-scale calculation,
with computationally accessible low-lying singlet states identified as
spurious charge-transfer states between the chromophore and a distant residue.\cite{Kim2023}
The \sing{1} state from TDA-TDDFT yielded a vertial excitation at 3.62~eV 
consistent with experimental observations,
where the broadband absorption by the neutral chromophore ranges from 330 to 450~nm
(2.76--3.75~eV).\cite{Heim1994} 
The QM/MM calculation predicted the same excitation to be 0.18~eV higher.

\begin{figure}[t]
	\centering
	\includegraphics[height=4cm,trim=3cm 0.9cm 0cm 0cm,clip]{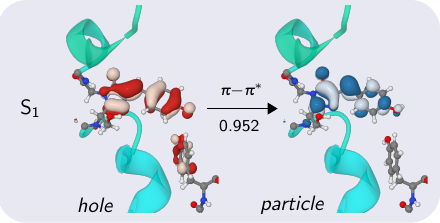}
	\caption{\normalfont
		Hole and particle wave functions as represented by the natural transition orbitals
		with the largest weight, shown in parentheses, from the $\omega$B97X calculation.
	}
	\label{fig:nto}
\end{figure}

The origin of the difference in the \sing{1} energies of the full-scale and QM/MM calculations
in this particular case can be inferred from the natural transition orbitals (NTOs).\cite{Martin2003}
Figure~\ref{fig:nto} shows the NTOs upon \sing{1} excitation, and 
for visual clarity, the chromophore and a neighboring residue are depicted.
Hole and particle wave functions of the \sing{1} state clearly indicate
that the \sing{1} state can be largely described as a local 
$\pi$--$\pi^\ast$ excitation within the chromophore.
Interestingly, a small participation of charge-transfer 
from the $\pi$-density of a nearby residue can be seen,
and presumably this has stabilized the \sing{1} state,
leading to the transition dipole moment calculated at the $\omega$B97X level
that is noticeably smaller than that of the QM/MM calculation (Table~\ref{tbl:calc}).
It is unclear how such excited-state details would affect the actual dynamics,
and the full implications remains to be determined,
but key long-range interactions that necessitate large-scale calculations 
can be captured in this approach.

\begin{figure}[t]
	\centering
	\includegraphics[height=3.5cm,trim=13cm 15cm 0cm 13cm,clip]{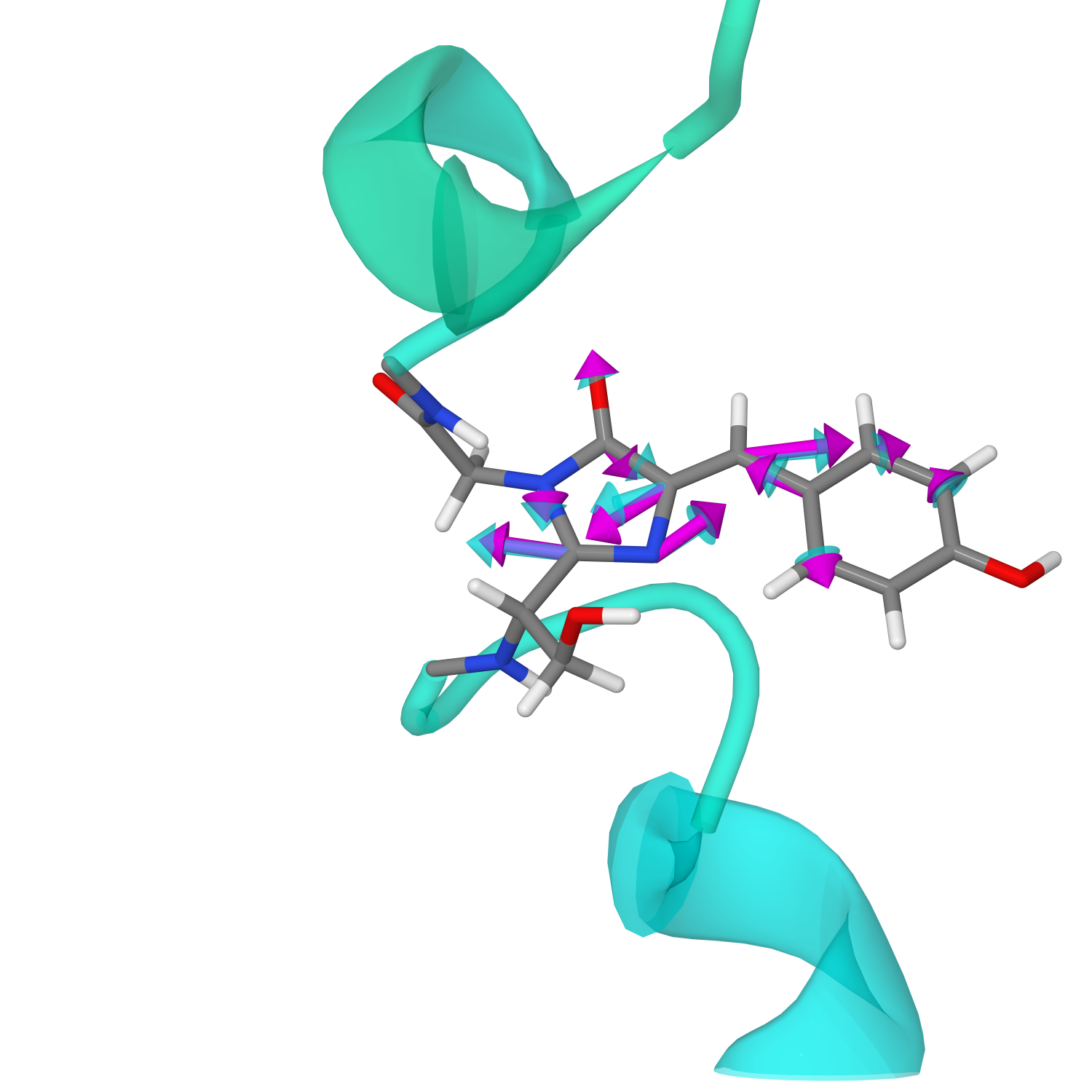}
	\caption{\normalfont
		\sing{1}--\sing{0} difference gradient ($\omega$B97X, magenta; ONIOM, cyan)
		with the arrows pointing to the negative sides.
		Vectors with norm greater than $10^{-2}$ au are shown.
	}
	\label{fig:grad}
\end{figure}

As excited-state dynamics to the first-order is governed 
by the nuclear forces on the potential energy surface,
we visualized in Figure~\ref{fig:grad}
the extracted \sing{1} nuclear gradient 
by computing the difference between excited- and ground-state gradients.
Since the employed geometry has not been optimized, 
after subtracting the ground-state portion from the total gradient (eq~\ref{eq:dE_mo}),
the resulting gradient reflects a change in the potential energy surface
upon the electronic excitation.

A noticeable difference between 
the nuclear gradients from $\omega$B97X and ONIOM calculations is observed
on the carbon atom on the imidazolone connected to the methine bridge,
suggesting the distant residue effect.
Indeed, similar visualizations with greater details showed
non-negligible participation of distant residues,
supporting the notion that excited-state dynamics 
can encompass the entire protein (\sifig{1}).
	Not unexpectedly, the electronic embedding in the ONIOM calculation
	also captured the nonvanishing \sing{1}--\sing{0} difference gradient
	outside the chromophore, but differing significantly from the $\omega$B97X result
	(\sifig{2}).
Focusing on relatively larger forces (i.e., the negative gradient), 
centered on the chromophore, the directions closely follow
the qualitative interpretation of the NTOs,
wherein the bonding-to-anti-bonding character change is reflected (Figure~\ref{fig:nto}).
For instance, 
the $\pi$-bonding orbital between the imidazolone and the methine bridge
becomes an anti-bonding orbital with depleted electron density,
suggesting a decrease in bond order and a bond elongation.
Fundamentally, the excited-state dynamics 
is governed by the total nuclear gradient encompassing the ground-state component,
which has been handily ignored in our interpretation.

\subsection{Multi-GPU Parallel Performance}

\begin{figure}[t]
	\centering
	\includegraphics[height=7cm,trim=1cm 0.2cm 0.3cm 0cm,clip]{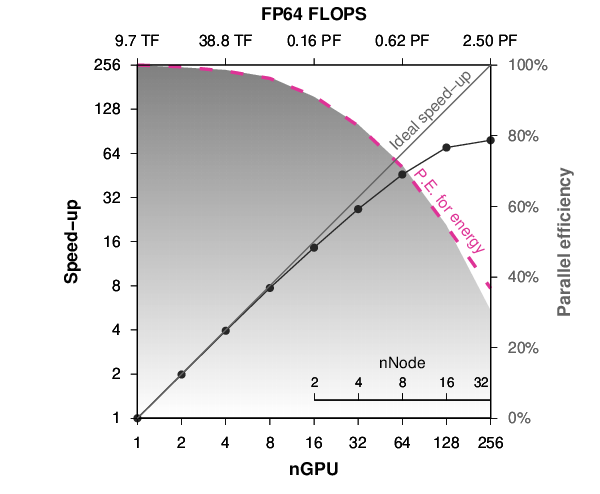}
	\caption{\normalfont
		Speed-up (black) and parallel efficiency (gray) of TDDFT gradient.
		The diagonal line indicates the ideal speed-up.
		Parallel efficiency of TDDFT energy calculation, 
		reproduced from ref~\citenum{Kim2023},
		is also plotted.
	}
	\label{fig:speed}
\end{figure}

Figure~\ref{fig:speed} shows the speed-up and parallel efficiency
as a function of the number of GPUs
for the TDDFT gradient calculation of the GFP system.
The parallel efficiency was measured by the ratio with respect to 
the 1-GPU wall time while doubling the number of GPUs.
Excellent parallel efficiency of $>$95\% is observed
within a single node comprising 8~GPUs.
Parallel efficiencies gradually decrease to 31\%
at 256 GPUs, and these numbers are very close to the parallel efficiency of 
the TDDFT energy calculation in our previous report,\cite{Kim2023}
suggesting that the same number of GPUs can be used for both energy and gradient calculations
without incurring computational imbalance.
	The relative timings of our implementation of TDDFT gradient
	with the a priori DFT and TDDFT energy calculations were
	$1:1.3:3.1$, respectively, further assuring that all the calculations yield
	comparable timings in the same order of magnitude (\sitab{3}).
In our largest computation with 256 A100 GPUs, 
the TDDFT gradient calculation took
a mere 1.2~h of time, empowering practical applications of TDDFT for very-large systems.

\begin{table}[t]
	\caption{Timings for the \sing{1} gradient calculation of GFP\textsuperscript{\emph{a}}}
	\label{tbl:time}
	\begin{ruledtabular}
		\begin{tabular}{lccccc}
			\textit{n}GPU\textsuperscript{\emph{b}} & total (s)\textsuperscript{\emph{c}}
			& Z-vector (s) & $\nabla$ERI (s)& $\nabla$XC (s) 
			& $\frac{\nabla{\text{ERI}}}{\nabla{\text{XC}}}$ \\
			\hline 
			1   & 343\,643 (100.0\%)& 167\,503 &  112\,421 & 1\,453 & 77.4\\
			2   & 172\,857  (99.4\%)&  84\,239 &   56\,430 &    739 & 76.4\\ 
			4   &  87\,064  (98.7\%)&  42\,474 &   28\,243 &    384 & 73.6\\ 
			8   &  44\,457  (96.6\%)&  21\,667 &   14\,207 &    225 & 63.1\\ 
			16  &  23\,580  (91.1\%)&  11\,719 &    7\,191 &    133 & 54.2\\ 
			32  &  12\,894  (83.3\%)&   6\,571 &    3\,684 &     80 & 46.2\\ 
			64  &   7\,484  (71.7\%)&   3\,895 &    1\,935 &     59 & 33.0\\ 
			128 &   4\,903  (54.8\%)&   2\,716 &    1\,065 &     41 & 26.0\\
			256 &   4\,263  (31.5\%)&   2\,360 &       696 &     49 & 14.1\\
		\end{tabular}
	\end{ruledtabular}

	\raggedright
	\textsuperscript{\emph{a}} Measured at the TDA-TDDFT level 
	with $\omega$B97X functional and def2-SVP basis set.
	Calculations were restarted with the converged molecular orbitals and excitation amplitudes.
	\textsuperscript{\emph{b}} Nvidia A100-80GB GPUs were used.
	\textsuperscript{\emph{c}} Parallel efficiency with respect to 1-GPU utilization is given in parentheses.
\end{table}

Table~\ref{tbl:time} provides a detailed timing analysis of wall time
by decomposing it into individual components.
The ERI gradient exhibited an excellent performance,
while the performance of the XC gradient reached a saturation point
at approximately 128 GPUs.
Furthermore, at 1~GPU, the ratio of workloads 
between ERI gradient to XC gradient is 77.4, and it decreases to 14.1 at 256 GPUs.
In fact, with 256~GPUs, out of overall 49~s,
the XC gradient kernel merely takes 8.6~s
(2.3~s for intermediates, 6.3~s for the gradient),
implying that most of the computational cost is allotted to the GPU preparation.
In other words, the XC gradient quickly reaches a plateau
due to the relatively short execution time of the GPU kernel, and
almost all times are spent on data management between devices,
leading to strong scaling stagnation almost immediately.
Better data management is thus important for further code optimization.
On the other hand, the ERI gradient has the most workloads 
in the entire gradient calculation,
and our load balancing scheme showed better scalability.
Clearly, the Z-vector portion remains approximately 50\% 
for all number of GPUs considered in this work,
rendering it the most time-consuming process.
The Z-vector iteration can be considered
as the direct matrix-vector product formation (eq~\ref{eq:sigma})
in the Davidson diagonalization
for TDDFT energies.\cite{Bauernschmitt1996,Leininger2001}

\begin{figure}[b]
	\centering
	\includegraphics[height=4cm,trim=0.3cm 0cm 0.3cm 0.2cm,clip]{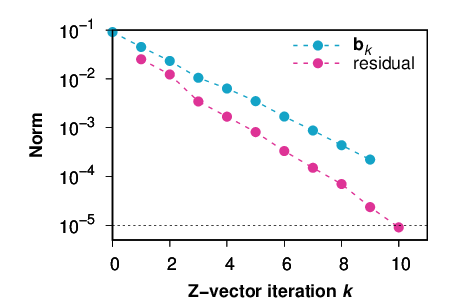}
	\caption{\normalfont
		Convergence profile of Z-vector iteration.
		Horizontal dashed line indicates the convergence threshold.
	}
	\label{fig:zconv}
\end{figure}

Figure~\ref{fig:zconv} shows the convergence profile of the Z-vector iterations
for solving the coupled-perturbed equations during the gradient evaluation.
Convergence was attained in 10 iterations, 
during which the residual norm decreased exponentially.
The norm of the additional subspace vector $\mathbf{b}_k$ also 
becomes exponentially smaller as the iterations proceeded.
As an undesirable side effect,
the product of the density matrix $\mathbf{B}_k$ (in eq~\ref{eq:zfock}) and the Schwarz upper bound of ERI,
which is used as the screening parameter, unintentionally neglect important ERI calculations in practice, potentially hampering convergence.
To mitigate such numerical instability 
at the expense of losing the benefit of efficent density screening,
we normalized $\mathbf{b}_k$ before the Fock build in eq~\ref{eq:sigma}, after which
it was back-converted to the original norm.
We suggest that 
a dynamical scheme to adjust the norm by comparing with previous iterations
can be helpful in reinstating density screening to accelerate the Z-vector iteration.

\begin{figure}[t]
	\centering
	\includegraphics[height=3.9cm,trim=0.cm 0.3cm 0.6cm 0.1cm,clip]{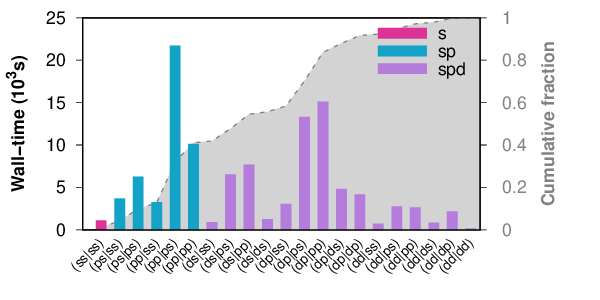}
	\caption{\normalfont
		Wall-times of GPU kernels for ERI derivatives measured with 1~GPU.
		ERIs are sorted in order of increasing the angular momentum.
		The cumulative fraction of time on each kernel is shown in gray.
	}
	\label{fig:eri}
\end{figure}

Figure~\ref{fig:eri} shows the time taken by all GPU kernels for ERI derivatives
grouped into three clusters based on the highest angular momentum before the differentiation.
Although approximately 45\% of the cost is attributed to the computation of
certain derivatives, $(pp|ps)$, $(dd|ps)$ and $(dp|pp)$
in this particular case of the excited-state of GFP,
this dependence is not only determined by how the basis functions are defined
in terms of the number of Gaussian functions in the given angular momentum 
and their contraction details, which determine the loop-counts in the code,
but also by the nature of the excited state.
The latter determines the pattern of nonzero elements
in the density matrix used to screen negligble ERIs prior to computation,
which is also an important factor.
Interestingly, adding a single $d$-polarization function to heavy atoms
as has been done in the employed def2-SVP basis set,
increased the total wall time for ERI derivatives by 2.4-fold.
This represents, in some respect, the necessary cost towards more realistic calculations.

\begin{figure}[t]
	\centering
	\includegraphics[height=5.3cm,trim=0.5cm 0.3cm 0.2cm 0.3cm,clip]{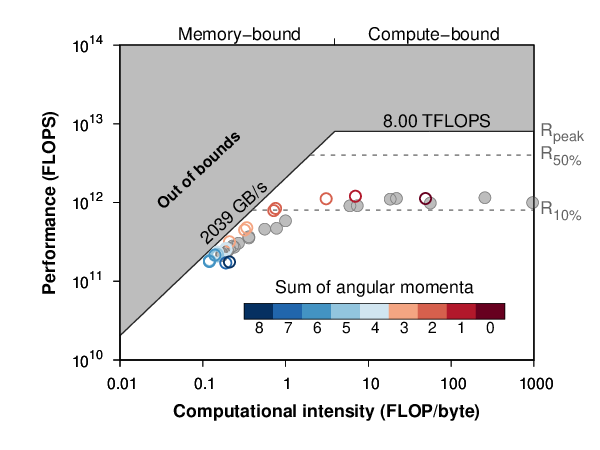}
	\caption{\normalfont
		Roofline analysis of ERI derivatives. 
		The peak performance was controlled by the employed profiler, Nsight Compute.
		Performance of ERI kernels, reproduced from ref~\citenum{Kim2023},
		are also plotted with gray circles.
	}
	\label{fig:roof}
\end{figure}

To provide deeper insights into the fundamental performance of these GPU kernels,
we performed a roofline analysis,\cite{Williams2009} as depicted in Figure~\ref{fig:roof}.
The GPU kernels for ERI derivatives are distributed across
the memory- and compute-bound regions,
which are bounded by the memory bandwidth and the peak performance, $R_\mathrm{peak}$.
As the involved angular momenta increase, denoted by their sum in the figure,
the kernels are shifted from the compute-bound to the memory-bound regions.

As seen in the compute-bound region in Figure~\ref{fig:roof}, the GPU kernels of ERI derivatives
utilize approximately 15\% of $R_\mathrm{peak}$, in double precision (FP64).
It is worth nothing that $R_\mathrm{peak}$ assumes 
full utilization of fused-multiply-add operation,
where a multiplication and an addition are performed simultaneously,
and hence, we anticipate that further speed gains could be achieved
with more sophisticated code optimizations.
When compared to the ERI kernels, represented by gray circles,
all the kernels for ERI derivatives 
clearly shift towards the memory-bound region
because the operational intensity is dominated by
building larger intermediate integrals for the ERIs with a higher angular momentum,
which involves increased memory access.
The ERIs with higher angular momenta are mostly memory-bound for the same reason.

\section{Conclusions}

We presented a multi-GPU implementation of analytical nuclear gradient of TDA-TDDFT
using Gaussian-type basis sets with RSH functionals, 
with a focus on computing large-scale molecular systems.
We revisited the underlying theory and algorithms
to transform the integral direct procedure
into an efficiently computable form in a massively parallel GPU computing environment.
We demonstrated full-scale calculations of the excited-state nuclear gradient
of a realistic protein system consisting of 4353 atoms
at the $\omega$B97X/def2-SVP level of theory.
Achieving the parallel efficiency of $>$70\% with up to 64 GPUs and 31\% with 256 GPUs,
the parallel efficiency of the nuclear gradient calculations 
was comparable to the excited-state energy counterpart,
showcasing a well-balanced usage of a distributed resource up to 256 GPUs
with 2.5~peta-FLOPS of raw computing power.

Heterogeneous computing is becoming increasingly prevalent
as GPU-accelerated systems dominate the supercomputing landscape.\cite{Atchley2023,Chang2024}
Quantum chemistry codes that fully exploit the state-of-the-art hardwares,
as discussed in this work, will be extremely helpful 
in efficiently exploring the electronic structure of excited states 
for very large molecular systems.
This will offer an exciting simulation platform for 
advancements in chemical and materials science.



\section*{Acknowledgements}
Part of this work was carried out during the stay of I.K. at MIT as a visiting researcher.
T.V.V. and Y.M.R. acknowledge Samsung Electronics for financial support
(IO221227-04385-01 and IO211126-09175-01).
The authors thank Dr. Xin Li and Prof. Patrick Norman at KTH for helpful discussions.
I.K. thanks Mandy S. Kim for assistance with data collection.
Computational resources were provided by the Supercomputing Center of Samsung Electronics.

\section*{Author contributions}
I.K., T.V.V., Y.M.R., W.-J.S. and H.-J.K. conceived and designed the project and wrote the manuscript.
I.K. developed GPU algorithms and wrote the code.
D.J, L.W. and A.A. performed quantum chemistry calculations.
J.Y. assisted with parallel computing set-up.
S.K. and Y.C. aided in interpreting the results.
I.J., S.L. and D.S.K. organized the project and supervised the computational study.
All authors analyzed the data, discussed the results, 
and contributed to various portions of the manuscript.



\bibliographystyle{achemso}
\bibliography{refs}

\end{document}